 \documentclass[twocolumn]{aastex701}
\usepackage[T1]{fontenc}
\usepackage{ae,aecompl}
\usepackage{graphicx}
\usepackage{amsmath}
\usepackage{amssymb}
\usepackage{supertabular}
\usepackage{xcolor}
\usepackage{academicons}
\usepackage{multirow}
\usepackage{threeparttable} %\usepackage{longtable}
\usepackage[normalem]{ulem}  %  \sout 
\usepackage{times}
\usepackage{hyperref}% 

\begin{document}
%--------|---------|---------|---------|---------|---------|---------|---------|
\title{Discovery of the anti-glitch in PSR J1835$-$1106}

\author{Mingyang Wang}
\affiliation{Department of Astronomy, Xiamen University, Xiamen 361005, China}
\email{1751446586@qq.com}  

\author{Peng Liu}
\affiliation{Department of Astronomy, Xiamen University, Xiamen 361005, China}
\affiliation{Xinjiang Astronomical Observatory, Chinese Academy of Sciences, Urumqi 830011, China}
\email{34820210156871@stu.xmu.edu.cn}  

\author[0000-0002-5381-6498]{Jianping Yuan}
\affiliation{Xinjiang Astronomical Observatory, Chinese Academy of Sciences, Urumqi 830011, China}
\affiliation{State Key Laboratory of Radio Astronomy and Technology, Beijing 100101, China}
\affiliation{Xinjiang Key Laboratory of Radio Astrophysics, 150 Science 1-Street, Urumqi 830011, China}
\email{yuanjp@xao.ac.cn}  

\author[0000-0002-0786-7307]{Ang Li}%
\affiliation{Department of Astronomy, Xiamen University, Xiamen 361005, China}
\email[show]{liang@xmu.edu.cn}  

\author[0000-0003-3127-0110]{Youli Tuo}
\affiliation{Institut f\"{u}r Astronomie und Astrophysik, Kepler Center for Astro and Particle Physics, Eberhard Karls Universit\"{a}t Tübingen, Sand 1, 72076 T\"{u}bingen, Germany}
\email{youli.tuo@astro.uni-tuebingen.de}  

\author{Shijun Dang}
\affiliation{School of Physics and Electronic Science, Guizhou Normal University, Guiyang 550001, China}
\email{dangsj@gznu.edu.cn}  

\author{Weihua Wang}
\affiliation{College of Mathematics and Physics, Wenzhou University, Wenzhou 325035, China}
\email{wang-wh@pku.edu.cn}  

\author{Mingyu Ge}
\affiliation{Key Laboratory of Particle Astrophysics, Institute of High Energy Physics, Chinese Academy of Sciences, Beijing 100049, China}
\email{gemy@ihep.ac.cn}  

\author{Xia Zhou}
\affiliation{Xinjiang Astronomical Observatory, Chinese Academy of Sciences, Urumqi 830011, China}
\affiliation{State Key Laboratory of Radio Astronomy and Technology, Beijing 100101, China}
\affiliation{Xinjiang Key Laboratory of Radio Astrophysics, 150 Science 1-Street, Urumqi 830011, China}
\email{zhouxia@xao.ac.cn}  

\author{Na Wang}
\affiliation{Xinjiang Astronomical Observatory, Chinese Academy of Sciences, Urumqi 830011, China}
\affiliation{State Key Laboratory of Radio Astronomy and Technology, Beijing 100101, China}
\affiliation{Xinjiang Key Laboratory of Radio Astrophysics, 150 Science 1-Street, Urumqi 830011, China}
\email{na.wang@xao.ac.cn}  

\begin{abstract}
We report the detection of an anti-glitch with a fractional frequency change of $\Delta\nu/\nu=-3.46(6)\times10^{-9}$ in the rotation-powered pulsar PSR J1835$-$1106 at MJD 55813, based on timing observations collected with the Nanshan 26-m and Parkes 64-m radio telescopes from January 2000 to July 2022.
A comparison of the average pulse profiles within $\pm300$ d of the event reveals no significant morphological changes. 
We also estimate the angular velocity lag between the normal and superfluid components at the time of the glitch, showing that one of the superfluid glitch models
is incompatible with PSR J1835$-$1106 due to its insufficient spin-down rate and angular velocity lag.
The wind braking scenario offers a viable alternative, consistent with the observed spin-down behavior, glitch amplitude, and post-glitch recovery.
High-cadence, high-sensitivity monitoring of similar events is essential to distinguish between internal (superfluid) and external (wind-related) glitch mechanisms.
\end{abstract}

\keywords{
%Unified Astronomy Thesaurus concepts: 
%Compact objects (288); 
%Dark matter (353); 
%Gamma-ray bursts (629);
%Gravitational waves (678); 
%High energy astrophysics (739)
%Neutron star cores (1107); 
Neutron stars (1108);
Pulsars (1306)
%Radio pulsars(1353)
%Relativistic stars(1392)
}

\section{Introduction} \label{sec:intro}
Pulsars are rapidly rotating, highly magnetized neutron stars known for their exceptional rotational stability. However, some exhibit irregular spin behavior, most notably in the form of glitches~\citep{2022RPPh...85l6901A}. A glitch is characterized by a sudden, discrete increase in spin frequency occurring over a very short timescale. These events typically appear as an abrupt acceleration in the pulsar’s rotation, often followed by a gradual recovery phase \citep{2016MNRAS.460.1201H,2024MNRAS.532..859L,2022MNRAS.510.4049B,2010ATel.2889....1E}, likely involving internal processes such as the transfer of angular momentum from the neutron star's superfluid interior to its charged component. In contrast, a few rare events show a sudden decrease in spin frequency, known as anti-glitches (also named spin-down glitches)~\citep{2013Natur.497..591A,2020ApJ...896L..42Y,2014MNRAS.440.2916S,2019ApJ...879..130R}.
Previous studies have shown that anti-glitches predominantly occur in objects with very strong surface magnetic fields—magnetars—and are often accompanied by radiative outbursts and changes in pulse profile, indicating a close association with magnetospheric activity~\citep{2014ApJ...784...37D,2019AN....340..340H}.

Recently, \cite{2024ApJ...967L..13T} reported the radiatively quiet anti-glitch in the rotation-powered pulsar (RPP) PSR B0540$-$69, noting that its pulse profile remained essentially unchanged before and after the event. Subsequently, \cite{2024ApJ...977..243Z} identified a series of anti-glitch episodes in the RPP PSR J1522$-$5735; similarly, these events showed no detectable alterations in pulse shape. These observations suggest that anti-glitches in RPPs may originate from physical mechanisms fundamentally similar to those of normal glitches.

In the present work, we report a new anti-glitch in the RPP PSR J1835$-$1106, based on timing observations from the Nanshan 26-m and Parkes 64-m radio telescopes.
The paper is organized as follows: Section \ref{sec:obs} provides a brief overview of the timing observations conducted with these telescopes. In Section \ref{sec:res}, we analyze the anti-glitch, while Section \ref{sec:dis} explores potential triggering mechanisms. Finally, Section \ref{sec:con} summarizes the main findings of the study.

\begin{table*} 
\small
\centering
\caption{The parameters of PSR J1835$-$1106, where the number in parentheses represents the uncertainty of the last significant digit for each parameter. Columns 1--11 represent the pulsar name (J2000), right ascension (RA), declination (DEC), period ($P$), period derivative ($\dot{P}$), proper motion in RA (PMRA), proper motion in DEC (PMDEC), dispersion measure (DM), surface dipole magnetic field ($B_{\rm s} = 3.2 \times 10^{19} \sqrt{P\dot{P}} $), epoch and characteristic age ($\tau_{\rm c} = P/(2\dot{P})$), respectively.
\label{tab:pa}}  %%%%%%%%%%%%%%%%%%%% table 1 %%%%%%%%%%%%%%%%%%
    \setlength{\tabcolsep}{5.5pt}  
\begin{tabular}{lcccccccccc}
  \hline   \hline 
PSR &RA &DEC &$P$ &$\dot{P}$ &PMRA &PMDEC &DM &$B_{\rm s}$ &Epoch &Age  \\ 
 &(hh:mm:ss) &($\ast^{\degr}:\ast^{\arcmin}:\ast^{\arcsec}$) &(s) &(10$^{-14}$) 
&(mas/yr) &(mas/yr) &(cm$^{-3}$\,pc) &($10^{12}$ G) & MJD &(kyr) \\ 
\hline
J1835$-$1106  &18:35:18.28(27)$^{a}$ &$-$11:06:15(16)$^{a}$ &0.1659$^{a}$ &2.0544$^{a}$ &27(5)$^{b}$ &56(190)$^{b}$ & 132.8132$^{a}$ & 1.87 & 59400 & 128  \\  
\hline     \hline 
\end{tabular}
\begin{tablenotes}
\item[ ] \textit{Note}. References for parameters of the pulsar: 
$^a$ \citep{2024MNRAS.530.1581K}; 
$^b$ \citep{ZouHWM2005}.
\end{tablenotes} %\vspace{-0.2cm}
\end{table*}%-----------------------------------------------------

\section{Observations and Data analysis} \label{sec:obs}

% reset footnote counter so main-text footnotes start from 1
\setcounter{footnote}{0}

The Nanshan 26-m radio telescope is a fully steerable Cassegrain‐focus parabolic antenna with a primary dish diameter of 26 m. It is equipped with four cryogenic receivers covering five wavelength bands: 1.3 cm, 3.6 cm, 6 cm, 13 cm, and 18 cm. Initially, timing observations were performed with a 128‑channel analog filter bank (AFB), each channel having a bandwidth of 2.5 MHz. In January 2010, the system was upgraded to a 1024‑channel digital filter bank (DFB) centered at 1556 MHz, providing a total bandwidth of 512 MHz and an effective bandwidth of 320 MHz. Pulsar timing uses the 18 cm receiver at a center frequency of 1540 MHz with 320 MHz of usable bandwidth, covering the range 1381.25–1701.25 MHz~\citep{2001MNRAS.328..855W}. Observations are carried out at least three times per month, with each integration lasting 4 minutes prior to October 2017 and 8 minutes thereafter. It should be noted that no timing data were acquired between 2014 and 2016 owing to the receiver and backend upgrades.

The Parkes radio telescope is a fully steerable, 64 m diameter dish located in New South Wales, Australia, and is among the most prolific instruments for pulsar discovery. It hosts multiple receivers and digital backend systems; in this work, we utilize timing data obtained with the 13‑beam (20 cm) receiver and the Parkes Digital Filterbanks PDFB1, PDFB2, PDFB3 and PDFB4. These observations were conducted at a center frequency of 1369 MHz with a total bandwidth of 256 MHz, spanning 1241–1497 MHz~\citep{StaveleyWBD1996,HobbsMDJ2020}. The Parkes timing programme typically observes each target pulsar every 2–4 weeks, with total integration times ranging from 2 to 15 minutes per epoch.

Upon data acquisition, we processed the raw observations with PSRCHIVE~\citep{2004PASA...21..302H,2012AR&T....9..237V} to perform coherent dedispersion and fold subintegrated profiles into average pulse templates. To derive high‑precision pulse times-of-arrival (ToAs), we employed PSRADD to sum all folded profiles into a high–signal‑to‑noise standard template, then cross‑correlated each individual profile against this template to extract ToAs referenced to the observatory. Finally, these site‑ToAs were converted to the Solar System barycenter using TEMPO2~\citep{2006MNRAS.369..655H,2006MNRAS.372.1549E} together with the JPL DE440~\citep{2021AJ....161..105P} ephemeris and the Barycentric Coordinate Time (TCB) scale, thereby removing delays introduced by the Earth’s motion.

We used TEMPO2 to fit the ToAs data and obtain an accurate rotation phase model. The phase was expanded using a Taylor series \citep{2006MNRAS.372.1549E}:
\begin{equation}
\phi(t)
= \phi_0
+ \nu_0\,\bigl(t - t_0\bigr)
+ \frac{1}{2!}\,\dot{\nu}_0\,\bigl(t - t_0\bigr)^2
+ \frac{1}{3!}\,\ddot{\nu}_0\,\bigl(t - t_0\bigr)^3 \ .
%\tag{1-4}
\label{eq:1-4}
\end{equation}
Here, $t$ denotes the ToA, and $\phi_0$ , $\nu_0$ , $\dot{\nu}_0$ , $\ddot{\nu}_0$ are, at the reference epoch $t_0$, the pulse phase, spin frequency, and the first and second time derivatives of the spin frequency, respectively. The observed glitch was modeled by the following equation \footnote{The exponential recovery term is intentionally omitted here, as our timing analysis revealed no statistically significant recovery signature in the post-glitch data for PSR J1835$-$1106.} \citep{2006MNRAS.372.1549E}:
\begin{equation}
\phi_{\rm g}(t) = \Delta\phi + \Delta\nu_{\rm p}(t - t_{\rm g}) + \frac{1}{2!}\Delta\dot{\nu}_{\rm p}(t - t_{\rm g})^2 \ .
\label{eq:1-5}
\end{equation}
Here, $t_{\rm g}$ denotes the epoch at which the spin‑down glitch occurs, $\Delta\nu_{\rm p}$ and $\Delta\dot{\nu}_{\rm p}$ are the permanent increments in the spin frequency and its first derivative, respectively, and $\Delta\phi$ is the phase offset immediately following the glitch.

\section{Results} \label{sec:res}

\begin{table*}
\centering
\caption{Pre- and post-glitch timing solutions for PSR J1835$-$1106. 
} 
\label{tab:par_j1835}
\small
\setlength{\tabcolsep}{12pt}
\begin{tabular}{@{}l  c c c c c c@{}}
  \hline   \hline 
 & $\nu$ (Hz)      & $\dot{\nu}$ ($10^{-13}$\,s$^{-2}$) 
           & Epoch (MJD)   & ToA numbers         & RMS ($\mu$s)   & Data Span (MJD) \\ 
\hline
Pre-glitch 1        & $6.0273153061(3)$  & $-7.4916(4)$ 
           & 51895        & 74             & 4619          & 51499--52225    \\
Post-glitch 1      & $6.0271780300(3)$  & $-7.4840(2)$ 
           & 54019        & 366            & 67244         & 52226--55812    \\
Post-glitch 2       & $6.0269456679(2)$  & $-7.4818(7)$ 
           & 57614        & 200            & 25408         & 55814--59457    \\
\hline     \hline 
\end{tabular}
\end{table*}

PSR J1835$-$1106 was discovered in 1996 during the Parkes Southern Pulsar Survey~\citep{1996MNRAS.279.1235M} and has since been detected across multiple wavebands. As summarized in Table~\ref{tab:pa}, it has a spin period of approximately 166 ms~\citep{2024MNRAS.530.1581K}, a characteristic age of about 128 kyr, and an inferred surface dipole magnetic field of $\sim1.87 \times 10^{12}$\,G. 

Over the course of its observational history, PSR J1835$-$1106 has exhibited two previously known small glitches\footnote{\href{https://www.jb.man.ac.uk/~pulsar/glitches/gTable.html}{https://www.jb.man.ac.uk/~pulsar/glitches/gTable.html}}~\citep{2004MNRAS.354..811Z,EspinozaLSK2011,2013MNRAS.429..688Y}. The first, at MJD 52226, was reported by \citet{2004MNRAS.354..811Z}, while the second, at MJD 58488, is listed in the Jodrell Bank Glitch Catalogue.
Through analysis of timing residuals spanning January 2000 to July 2022, we identified both a previously-reported small normal glitch (glitch 1) and a new anti-glitch (glitch 2) in PSR J1835$-$1106. We refitted the parameters of the normal glitch at MJD 52226 but found no evidence of the event at MJD 58488 in our dataset. The timing models before and after the detected events are presented in Table~\ref{tab:par_j1835}, and the corresponding glitch parameters are listed in Table~\ref{tab:glitch}.
\begin{figure}%[htbp]
  \centering
  \includegraphics[width=0.5\textwidth]{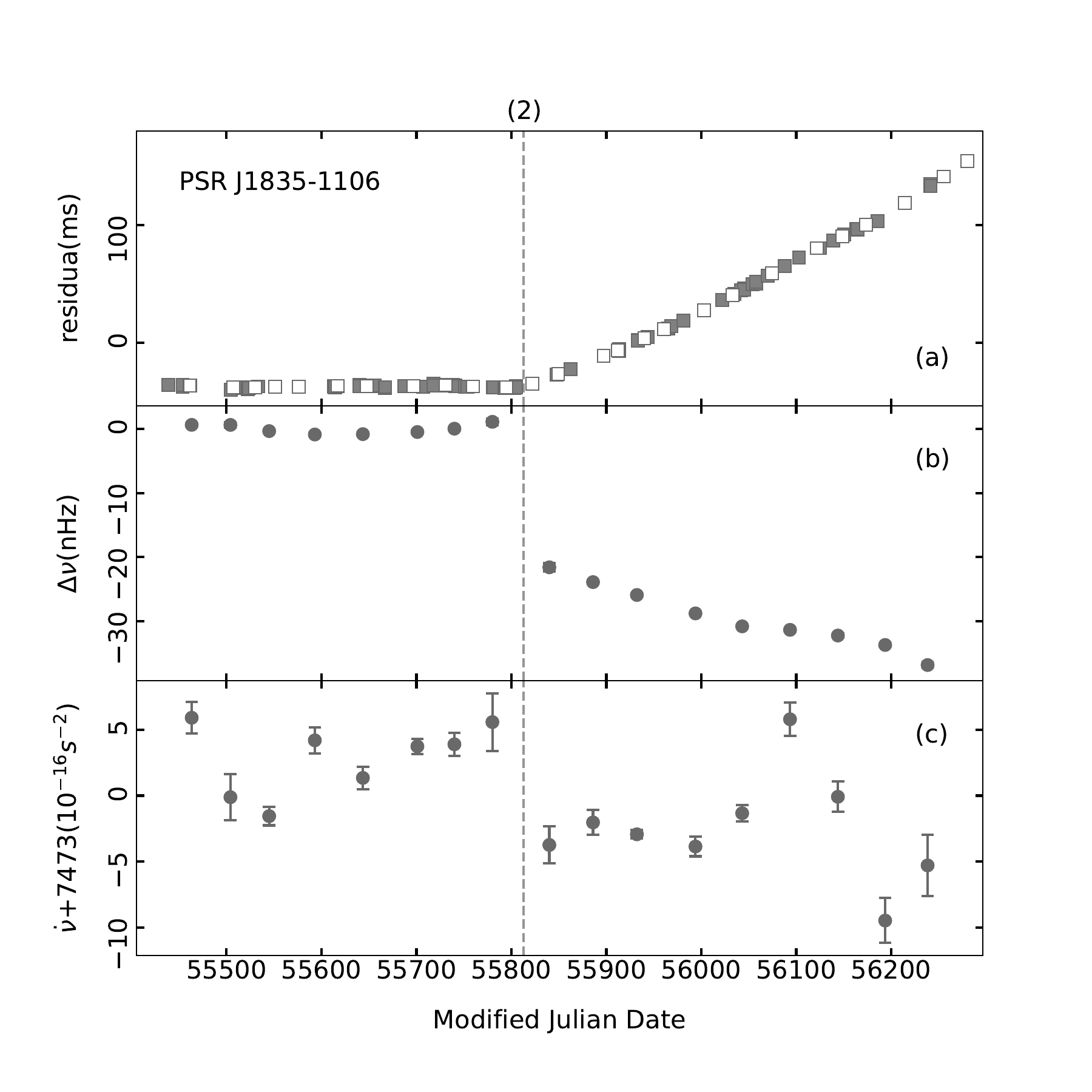} % 插入图片
  \caption{The anti-glitch of PSR J1835$-$1106. Grey dashed lines mark the epochs at which the glitches occur, with the event numbers indicated in parentheses above. Panel (a) shows the timing residuals, where filled squares represent data from the Nanshan 26 m telescope and open squares correspond to data from the Parkes 64 m telescope. Panel (b) displays the relative change in spin frequency, defined as the difference between the observed spin frequency and that predicted by the pre‑glitch timing model. Panel (c) presents the temporal evolution of the spin‑frequency derivative.} 
  \label{fig:1835_g2} 
\end{figure}

\begin{table} 
%\small
\footnotesize
\centering
\caption{Glitch parameter fitting for PSR J1835$-$1106. Glitch 2 corresponds to the newly identified anti-glitch event.
\label{tab:glitch}}    %%%%%%%%%%%%%%%%%%%% table 3 %%%%%%%%%%%%%%%%%%
    \setlength{\tabcolsep}{5pt}  
\begin{tabular}{lccc}
 \hline \hline
 Parameter             &Glitch 1$^a$   &Glitch 2$^b$ &Glitch 3$^c$\\
 \hline
 Glitch epoch (MJD)    &52226(9) &55813(9) &58488(2)\\
 $\Delta\nu/\nu$ ($10^{-9}$) &18.23(19)   &$-$3.46(6) &9.3(6)\\
 $\Delta\dot{\nu}/\dot{\nu}$ ($10^{-3}$) &$-$0.81(11) &0.47(3) &4(1)\\
RMS timing residual ($\mu$s) & 730 & 942 &$-$\\
Data span (MJD)       & 51924--52458 & 55377--56326 &$-$ \\
      \hline \hline
\end{tabular}
\begin{tablenotes}
\item[ ] \textit{Note}. Glitch reported by $^a$\cite{2004MNRAS.354..811Z};
$^b$Present work;
$^c$Jodrell Bank glitch catalogue. 
\end{tablenotes} %\vspace{-0.2cm}
\end{table}%-----------------------------------------------------

\begin{figure}%[htbp]
  \centering
  \includegraphics[width=0.5\textwidth]{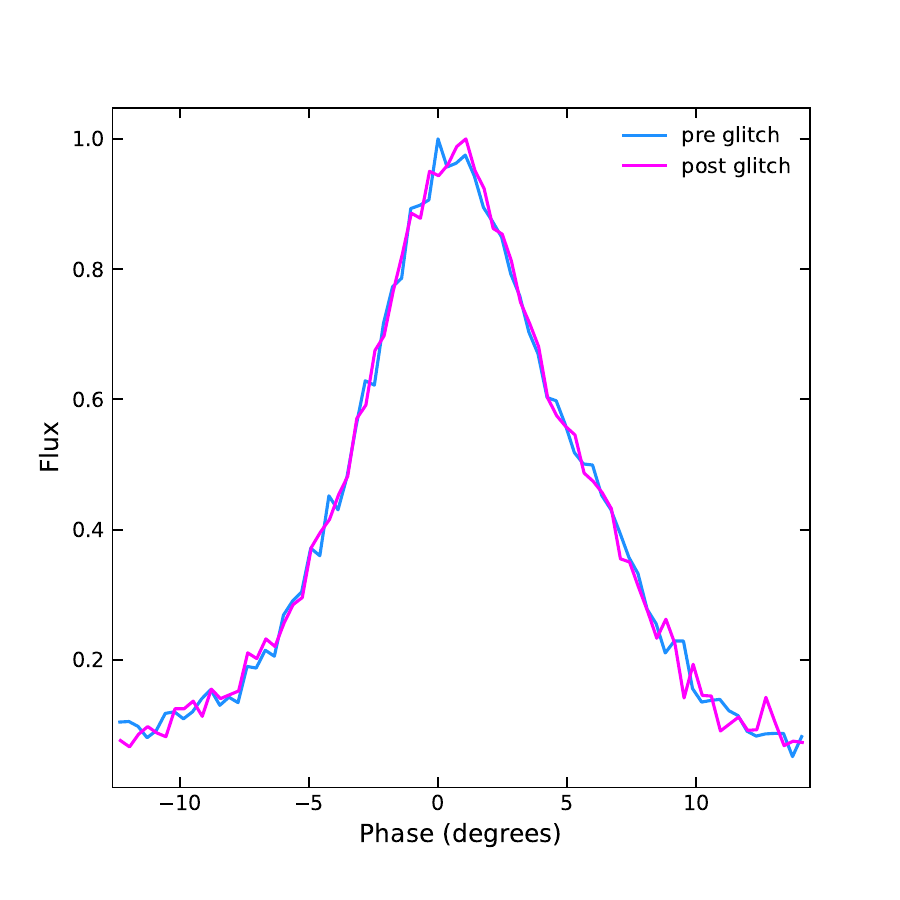} % 插入图片
  \caption{Average pulse profiles of PSR J1835$-$1106 in the $\pm$300 d interval surrounding the anti-glitch.} 
  \label{fig:profile} 
\end{figure}

The newly detected anti-glitch occurred at MJD 55813(9), with a fractional frequency change of $\Delta\nu/\nu = -3.46(6) \times 10^{-9}$ (see Fig.~\ref{fig:1835_g2}). We also examined the average pulse profiles within a $\pm$300-day window around this event (Fig.~\ref{fig:profile}) and found no significant changes in pulse morphology before or after the glitch.
As noted earlier, many anti-glitches in magnetars are accompanied by magnetospheric activity. However, the anti-glitch reported here occurred in an RPP, with no detectable changes in pulse profile. 
It should be emphasized, however, that the absence of detectable changes in the radio band does not rule out possible emission variations in beam components outside our line of sight or at other energies.
Unfortunately, PSR J1835$-$1106 is extremely faint in the $\gamma$-ray band, and our current $\gamma$-ray observations lack the sensitivity to detect any flux changes. 
Using the charge-density variation expressions from \citet{2022MNRAS.513.5861S} and \citet{2024MNRAS.528.7458B}, we relate the change in spin-down rate ($\dot \nu$) to the change in magnetospheric charge density $\Delta\rho$ and estimate \(\Delta\rho/\rho_{\rm GJ}\approx3.08\times10^{-4}\). where $\rho_{\rm GJ}$ is the Goldreich–Julian charge density~\citep{1969ApJ...157..869G}.
This value is about two orders of magnitude smaller than those reported in \citet{2022MNRAS.513.5861S}, implying that any associated radiative variation would be expected to be vanishingly weak.

\section{Discussion} \label{sec:dis}

Previously, the anti-glitch in RPP PSR J1522$-$5735 was interpreted as inward motion of vortex lines driven by external forces, leading to angular momentum transfer from the crust to the superfluid~\citep{2024ApJ...977..243Z}. In contrast, the anti-glitch in RPP PSR B0540$-$69 was explained via the $\Delta V$ effect proposed by \cite{2014ApJ...797L...4K}, where partial conversion of superfluid to normal matter increases the moment of inertia of the normal component while reducing that of the superfluid component. 

According to the modified vortex creep model~\citep{2014ApJ...797L...4K}, the spin frequency change during an anti-glitch is expressed as:
\begin{equation}
\Delta \Omega_{\rm{c}} = - \frac{I_{\rm{s0}} - I'_{\rm{c}} \Delta \Omega_0}{I_{\rm{c0}} + I'_{\rm{s}} \Delta \Omega_0} \Delta \Omega_{\rm{s}} \ ,
\label{eq:delta_omega}
\end{equation}
where $I_{\rm{s0}}$ and $I_{\rm{c0}}$ are the moments of inertia of the superfluid and normal components before the glitch, $\Delta \Omega_0$ is the angular velocity lag between the components prior to the event, and $I'_{\rm{c}}=-dI_{\rm{s}}/d(\Delta \Omega)$. An anti-glitch is expected to occur when $I_{\rm{s0}} < I'_{\rm{c}} \Delta \Omega_0$. However, this mechanism requires stringent conditions: the superfluid density must exhibit significant sensitivity to $\Delta \Omega$, necessitating high internal temperatures ($T \gtrsim 10^7\,\mathrm{K}$) and a substantial angular velocity lag ($\Delta \Omega \gtrsim 1\ \mathrm{rad\ s^{-1}}$), which are likely met only in a few young pulsars.

Notably, PSR J1835$-$1106 exhibited a small glitch preceding its anti-glitch. If this precursor originated from vortex unpinning, it could allow estimation of whether the angular velocity lag $\Delta \Omega$ at the anti-glitch epoch (MJD 55813(9)) satisfies $\Delta \Omega \gtrsim 1\ \mathrm{rad\ s^{-1}}$. According to \citet{2024ApJ...967L..13T}, a normal glitch also preceded the anti-glitch in PSR B0540$-$69. Assuming the superfluid angular velocity remained unchanged after the normal glitch while the crust was braked, the inferred lag for PSR B0540$-$69 at the anti-glitch was $0.88\ \mathrm{rad\ s^{-1}}$, close to the critical threshold. However, applying the same method to PSR J1835$-$1106 yields $1.4\times10^{-3}\ \mathrm{rad\ s^{-1}}$, far below the required value. For PSR J1522$-$5735, the short interval between events ($1000\ \mathrm{d}$) also precludes sufficient time to accumulate a significant lag. Thus, if the precursor glitch in PSR J1835$-$1106 involved vortex unpinning, the subsequent anti-glitch cannot be explained by the $\Delta V$ effect.

The estimated spin lag of $1.4 \times 10^{-3}\ \mathrm{rad\ s^{-1}}$ is based on the assumption that the superfluid reservoir was completely emptied during the previous glitch. 
While this assumption may be overly simplistic, we argue that the conclusion—that the anti-glitch cannot be attributed to the $\Delta V$ effect—remains robust.
First, the superfluid reservoir in PSR J1835$-$1106 may not have been emptied completely. 
A statistical relationship between glitch activity and spin-down rate, $\dot\nu_{\rm g} = \sum \Delta \nu / (T_{\rm span}) \approx 0.01 |\dot\nu|$, has been observed in pulsars with $-13.5 < \log |\dot\nu| < -10.5$ \citep{2017A&A...608A.131F} or $-14< \log |\dot\nu| < -10.5$~\citep{2022MNRAS.510.4049B}, where $\sum \Delta \nu$ is the cumulative frequency increase due to glitches over the total observed interval $T_{\rm span}$.
For PSR J1835$-$1106, its spin down rate falls within the range $-13.5<{\rm log}|\dot\nu|<-10.5$. If $\dot\nu_{\rm g}=0.01|\dot\nu|$ applies to PSR J1835$-$1106, the expected change in frequency within the longest data span (from MJD 51499 to 59457 according to Table \ref{tab:par_j1835}) should be $\sum \Delta\nu=0.01|\dot\nu|\times 7958~{\rm d}=5.14\times 10^{-6}~{\rm Hz}$. However, the total observed change in frequency is $\sum \Delta\nu=(18.23+9.3)\times10^{-9}\nu=1.66\times 10^{-7}~{\rm Hz}$ according to Table \ref{tab:glitch}, smaller than what is expected by a factor of 32. This suggests that most of its superfluid angular momentum is either still stored in the reservoir or has been transferred by other mechanisms. 
In this case, the previously estimated spin lag $1.4\times 10^{-3}~{\rm rad~s^{-1}}$ should be regarded as a lower limit. 
Second, assuming no glitches occur and the spin lag accumulates steadily,
achieving $\Delta \Omega \gtrsim 1\ \mathrm{rad\ s^{-1}}$ would require both a large spin-down rate and prolonged intervals without glitches. PSR B0540$-$69, with its substantial spin-down rate ($\sim2.5\times10^{-10}\,\mathrm{s^{-2}}$) attributed to a pulsar wind nebula \citep{2007ApJ...662..988P} and a $\sim 7000$-day glitch-free interval, could accumulate such a lag. In contrast, PSR J1835$-$1106 would require approximately $~6800\ \mathrm{yr}$ under crustal motion assumptions to build up such a comparable lag. 
This strongly disfavors the $\Delta V$ effect as the origin of its anti-glitch.

Another noteworthy aspect is that the epoch uncertainty of this anti-glitch reaches 18 d. Under this relatively long timescale, some relatively slow physical mechanisms may also cause the occurrence of the anti-glitch. \cite{2013ApJ...768..144T} proposed a wind braking scenario that provides a possible explanatory framework. This model has been used to explain anti-glitches in magnetars such as 1E 2259$+$586 \citep{2014ApJ...784...86T} and may allow anti-glitches to occur with extremely small flux increases that remain undetected. Within the wind braking scenario, the anti-glitch is not an instantaneous event but accumulates from a phase of enhanced spin-down. When observing cadence is insufficient, this sustained spin-down manifests as a net spin frequency decrease in timing data. We assume the whole spin-down energy loss is going into the particle wind, according to this model, the rotational energy loss rate due to the particle wind is~\citep{2013ApJ...768..144T}:
\begin{equation}
\dot{E}_{\rm{w}} = \sqrt{\dot{E}_{\rm{d}}} \sqrt{\rm L_{p}} \ ,
\end{equation}
where $\dot{E}_{\rm d}$ is the magnetic dipole rotational energy loss rate and $L_{\rm p}$ is the particle wind luminosity.The epoch uncertainty allows enhanced spin-down to persist for up to $\Delta t \approx 18~\text{d}$. The extra rotational energy loss during this interval is:
\begin{equation}
\begin{split}
\Delta E_{\rm g} 
&= \left( \sqrt{\dot{E}_{\rm d}} \sqrt{L_{\rm p,\text{glitch}}} - \sqrt{\dot{E}_{\rm d}} \sqrt{L_{\rm p,\text{pre}}} \right) \Delta t \\
&= \sqrt{\dot{E}_{\rm d}} \sqrt{L_{\rm p,\text{pre}}} \left( \frac{ \sqrt{L_{\rm p,\text{glitch}}} }{ \sqrt{L_{\rm p,\text{pre}}} } - 1 \right) \Delta t \\
&= \dot{E}_{\text{pre}} ( \beta - 1 ) \Delta t \ ,
\end{split}
\end{equation}
where $L_{\rm p,\text{glitch}}$ denotes the average wind luminosity during the anti-glitch, $L_{\rm p,\text{pre}}$ denotes the average wind luminosity before the anti-glitch, $\dot{E}_{\rm pre} = 4 \pi^{2} I \nu | \dot{\nu} |$ is the pre-glitch energy loss rate, and $\beta = \sqrt{L_{p,\text{glitch}}} / \sqrt{L_{p,\text{pre}}}$. The energy loss $\Delta E_{\rm g}$ is observationally $\Delta E_{\rm g} = 4 \pi^{2} I \nu | \Delta \nu |$. Thus we obtain:
\begin{equation}
\beta = \frac{ \Delta E_{\rm g} }{ \dot{E}_{\rm pre} \Delta t } + 1 = \frac{ | \Delta \nu | }{ | \dot{\nu} | \Delta t } + 1 \approx 1.018 \ .
\end{equation}
This shows that due to PSR J1835$-$1106's small anti-glitch amplitude ($\Delta\nu/\nu = -3.46(9)\times10^{-9}$) and potentially long-duration wind activity (days to weeks), the luminosity increase during the event is only $\sim$3.6\% ($\beta^2 \approx 1.036$). Applying the same methodology to PSR B0540$-$69 and PSR J1522$-$5735 yields $\beta^2 \approx 1.001$ ($|\Delta \nu| \approx 1 \times 10^{-7}$ Hz, $\Delta t \approx 10$ d) and $\beta^2 \approx 1.080$ ($|\Delta \nu| \approx 5 \times 10^{-8}$ Hz, $\Delta t \approx 10$ d), respectively. %Such minor increases may not produce detectable radiative events. 
Employing the Parkes signal-to-noise ratio ($\mathrm{S/N}$) calculation (Eq.(1) in \citealt{2006MNRAS.372..777L}) to estimate the upper-limit $\mathrm{S/N}$ for a single observation of PSR~J1835$-$1106, yielding $\mathrm{S/N}\approx84$. 
Under these conditions, a 3.6\% flux enhancement in the case of PSR J1835$-$1106 would yield $\Delta\mathrm{S/N}\approx3.0$, i.e.\ right at the 3$\sigma$ detection threshold. 
In practice, however, most Parkes pulse-profile observations of this pulsar have $\mathrm{S/N}$ values between 20 and 40. Adopting a representative value of $\mathrm{S/N}= 30$ gives $\Delta\mathrm{S/N}\approx1.1$, making it highly unlikely that such a small flux increase could be detected during the glitch period with existing instruments. 
We mention here that these estimates strongly depend on the assumed duration of wind enhancement $\Delta t$, where using the epoch uncertainty as $\Delta t$ only provides an upper limit—the actual wind-enhancement timescale could be substantially shorter. Additionally, our calculation adopts time-averaged luminosity during the wind-enhancement phase, while instantaneous values could be significantly higher. Considering these factors, detectable radiative changes during anti-glitches in RPPs remain possible within the wind braking scenario. For example, assuming a wind duration of 10\% the epoch uncertainty for PSR J1835$-$1106 ($\Delta t \approx 1.8$ d) gives $\beta^2 \approx 1.39$. Therefore, future high-cadence observations of anti-glitches, searching for radiative state changes and determining whether spin-frequency variations are truly instantaneous, could help distinguish between wind braking and superfluid origins of anti-glitches.
Furthermore, within the wind braking scenario, the slight post-glitch increase in spin-down rate ($\Delta\dot{\nu}/\dot{\nu} = 0.47(3) \times 10^{-3}$) can be naturally explained by residual wind activity persisting after the anti-glitch.

%%%%%%%%%%%%%%%%%%%%%%%%%%%%%%%%%%%%%%%%%%%%%%%%%%%%%%%%%%%%%%%%%%%%%%%%%%%%
\section{Summary and conclusions} \label{sec:con}
In this study, we analyzed timing observations of PSR J1835$-$1106 from the Nanshan 26-m and Parkes 64-m radio telescopes and identified an anti-glitch at MJD 55813(9) with fractional size $\Delta\nu/\nu=-3.46(6)\times10^{-9}$, representing the third such event detected in a RPP. We compared the average pulse profiles within $\pm300$ d of the anti-glitch and found no significant morphological changes. %\sout{implying that the origin of this event is unlikely to be magnetospheric. }
This significantly differs from most previously observed anti-glitches. 

The anti-glitches observed in RPPs have traditionally been interpreted through different physical mechanisms. For instance, PSR J1522$-$5735's anti-glitch was attributed to inward vortex motion transferring angular momentum from the crust to the superfluid, while that of PSR B0540$-$69 was explained via the $\Delta V$ effect, where the moment of inertia shifts due to partial conversion of superfluid to normal matter. This latter mechanism requires a strong sensitivity of superfluid density to angular velocity lag, a condition likely only met in young pulsars with high internal temperatures and substantial angular velocity lag.

For PSR J1835$-$1106, the observed anti-glitch is preceded by a normal glitch, prompting investigation of whether the $\Delta V$ mechanism is applicable. However, estimated angular velocity lag ($\Delta \Omega \sim 1.4\times10^{-3}\ \mathrm{rad\,s^{-1}}$) is far below the required threshold of 1 rad/s. Even considering this value as a lower limit, the timescale required to accumulate the necessary lag in PSR J1835$-$1106—about 6800 yr—makes the $\Delta V$ effect implausible. This stands in contrast to PSR B0540$-$69, where the combination of a high spin-down rate and a long glitch-free interval makes significant lag buildup more feasible.

Given this, alternative explanations such as the wind braking model are considered. This model posits that anti-glitches result from temporary increases in particle wind luminosity, enhancing the pulsar’s braking torque and mimicking an anti-glitch in timing data. Under this framework, the anti-glitch is not a sudden event but a phase of enhanced spin-down spread over a period (up to the 18-day uncertainty in glitch epoch). Calculations for PSR J1835$-$1106 show that only a modest increase ($\sim3.6\%$) in particle wind luminosity could account for the observed glitch, consistent with the lack of strong radiative signals. Similar small luminosity increases are inferred for other RPPs, such as PSR B0540$-$69 and PSR J1522$-$5735.

However, the inferred luminosity increase is sensitive to the assumed duration of wind activity. If the actual duration is shorter than the epoch uncertainty (e.g., $10\%$ of 18 d), the instantaneous luminosity could be significantly higher (e.g., $\beta^2 \sim 1.39$), possibly producing detectable emission. Therefore, high-cadence observational campaigns during suspected anti-glitch events are essential to test for associated radiative changes and to distinguish between superfluid and wind-related origins.

Note that the persistent post-glitch increase in spin-down rate in PSR J1835$-$1106 can also be naturally interpreted as residual wind activity within the wind braking scenario. Additionally, an evaluation of expected versus observed cumulative frequency changes indicates that much of the superfluid angular momentum remains stored or is released via alternative mechanisms, reinforcing the argument against the $\Delta V$ explanation for this pulsar.

In future studies, detailed investigations into the angular momentum exchange mechanisms between the superfluid interior and the normal component are crucial for building a self-consistent physical model of anti-glitches. Simultaneous analyses incorporating polarization measurements and other diagnostics may also help elucidate the connection between anti-glitches and pulsar emission, offering insights into whether their origin lies in magnetospheric processes or internal superfluid dynamics. Future high-cadence, high signal-to-noise, multi-frequency observations—such as those anticipated from the enhanced X-ray Timing and Polarimetry mission (eXTP; \citealt{2025arXiv250608101Z,2025arXiv250608369G,2025arXiv250608104L})—are expected to expand the sample of anti-glitches in RPPs and may reveal universal features underlying these enigmatic events.

\section*{Acknowledgments}
The work is supported by the National SKA Program of China (No.~2020SKA0120300), the Strategic Priority Research Program of the Chinese Academy of Sciences (No. XDB0550300), the National Natural Science Foundation of China (Nos. 12273028, 12494572 and~12041304), and XMU Training Program of Innovation and Enterpreneurship for Undergraduates.
The Nanshan 26-m Radio Telescope is partly supported by the Operation, Maintenance and Upgrading Fund for Astronomical Telescopes and Facility Instruments, budgeted from the Ministry of Finance of China (MOF) and administrated by the Chinese Academy of Sciences (CAS).
The Parkes radio telescope is part of the Australia Telescope National Facility which is funded by the Commonwealth of Australia for operation as a National Facility managed by CSIRO. This paper includes archived data obtained through the CSIRO Data Access Portal.

\vspace{5mm}
\facilities{Nanshan 26-m radio telescope, Parkes 64-m radio telescope}

\software{\texttt{TEMPO2} \citep{2006MNRAS.372.1549E,2006MNRAS.369..655H}, 
          \texttt{PSRCHIVE} \citep{2004PASA...21..302H,2012AR&T....9..237V}, 
          }

\bibliography{J1835}{}
\bibliographystyle{aasjournal}

\end{document}